\documentclass[longauth]{aa} 
\usepackage[utf8]{inputenc}
\usepackage{graphicx}
\usepackage[T1]{fontenc}
\usepackage{amsmath}
\usepackage{graphicx,overpic}
\usepackage{fixltx2e}
\usepackage{placeins}

\usepackage{txfonts}
\usepackage{natbib} 
\bibpunct{(}{)}{;}{a}{}{,} 

\usepackage{dashbox,framed,color}

\begin{document} 
	 
\title{Imaging Jupiter's radiation belts down to 127 MHz with LOFAR}

\author{
J.~N.~Girard\inst{37,38,1}\and 
P.~Zarka\inst{2}\and 
C.~Tasse\inst{3}\and 
S.~Hess\inst{4}\and 
I.~de Pater\inst{5}\and 
D.~Santos-Costa\inst{6}\and 
Q.~Nenon\inst{4}\and 
A.~Sicard\inst{4}\and 
S.~Bourdarie\inst{4}\and 
J.~Anderson\inst{7}\and 
A.~Asgekar\inst{8,9}\and 
M.~E.~Bell\inst{10}\and 
I.~van Bemmel\inst{8,11}\and 
M.~J.~Bentum\inst{8,12}\and 
G.~Bernardi\inst{13}\and 
P.~Best\inst{14}\and 
A.~Bonafede\inst{15}\and 
F.~Breitling\inst{16}\and 
R.~P.~Breton\inst{17}\and 
J.~W.~Broderick\inst{18}\and 
W.~N.~Brouw\inst{8,19}\and 
M.~Br\"uggen\inst{15}\and 
B.~Ciardi\inst{20}\and
S.~Corbel\inst{1}\and
A.~Corstanje\inst{21}\and 
F.~de Gasperin\inst{15}\and 
E.~de Geus\inst{8,22}\and 
A.~Deller\inst{8}\and 
S.~Duscha\inst{8}\and 
J.~Eisl\"offel\inst{23}\and 
H.~Falcke\inst{21,8}\and 
W.~Frieswijk\inst{8}\and 
M.~A.~Garrett\inst{8,24}\and 
J.~Grie\ss{}meier\inst{25,26}\and 
A.~W.~Gunst\inst{8}\and 
J.~W.~T.~Hessels\inst{8,27}\and 
M.~Hoeft\inst{23}\and 
J.~H\"orandel\inst{21}\and 
M.~Iacobelli\inst{8}\and 
E.~Juette\inst{28}\and 
V.~I.~Kondratiev\inst{8,29}\and 
M.~Kuniyoshi\inst{30}\and 
G.~Kuper\inst{8}\and 
J.~van Leeuwen\inst{8,27}\and 
M.~Loose\inst{8}\and 
P.~Maat\inst{8}\and 
G.~Mann\inst{16}\and 
S.~Markoff\inst{27}\and 
R.~McFadden\inst{8}\and 
D.~McKay-Bukowski\inst{31,32}\and 
J.~Moldon\inst{8}\and 
H.~Munk\inst{8}\and 
A.~Nelles\inst{21}\and 
M.~J.~Norden\inst{8}\and 
E.~Orru\inst{8}\and 
H.~Paas\inst{33}\and 
M.~Pandey-Pommier\inst{34}\and 
R.~Pizzo\inst{8}\and 
A.~G.~Polatidis\inst{8}\and 
W.~Reich\inst{35}\and 
H.~R\"ottgering\inst{24}\and 
A.~ Rowlinson\inst{10}\and 
D.~Schwarz\inst{36}\and 
O.~Smirnov\inst{37,38}\and 
M.~Steinmetz\inst{16}\and 
J.~Swinbank\inst{27}\and 
M.~Tagger\inst{25}\and 
S.~Thoudam\inst{21}\and 
M.~C.~Toribio\inst{39}\and 
R.~Vermeulen\inst{8}\and 
C.~Vocks\inst{16}\and 
R.~J.~van Weeren\inst{13}\and 
R.~A.~M.~J.~Wijers\inst{27}\and 
O.~Wucknitz\inst{35}
}
\institute{
AIM, UMR CEA-CNRS-Paris 7, Irfu, Service d'Astrophysique, CEA Saclay, F-91191 GIF-SUR-YVETTE CEDEX, France \and
LESIA, UMR CNRS 8109, Observatoire de Paris, 92195   Meudon, France \and
GEPI, Observatoire de Paris, CNRS, 5 place Jules Janssen, 92195 Meudon, France \and
ONERA, DESP, 2 Av. Edouard Belin, 31055 Toulouse, France \and
University of California, Department of Astronomy, 501 Campbell Hall, Berkeley CA 94720, USA \and
Southwest Research Institute, San Antonio, Texas, USA \and
Helmholtz-Zentrum Potsdam, DeutschesGeoForschungsZentrum GFZ, Department 1: Geodesy and Remote Sensing, Telegrafenberg, A17, 14473 Potsdam, Germany \and
ASTRON,the Netherlands Institute for Radio Astronomy, Postbus 2, 7990 AA Dwingeloo, The Netherlands \and
Shell Technology Center, Bangalore, India \and
CSIRO Australia Telescope National Facility, PO Box 76, Epping NSW 1710, Australia \and
Joint Institute for VLBI in Europe, Dwingeloo, Postbus 2, 7990 AA The Netherlands \and
University of Twente, The Netherlands \and
Harvard-Smithsonian Center for Astrophysics, 60 Garden Street, Cambridge, MA 02138, USA \and
Institute for Astronomy, University of Edinburgh, Royal Observatory of Edinburgh, Blackford Hill, Edinburgh EH9 3HJ, UK \and
University of Hamburg, Gojenbergsweg 112, 21029 Hamburg, Germany \and
Leibniz-Institut f\"{u}r Astrophysik Potsdam (AIP), An der Sternwarte 16, 14482 Potsdam, Germany \and
Jodrell Bank Center for Astrophysics, School of Physics and Astronomy, The University of Manchester, Manchester M13 9PL,UK \and
School of Physics and Astronomy, University of Southampton, Southampton, SO17 1BJ, UK \and
Kapteyn Astronomical Institute, PO Box 800, 9700 AV Groningen, The Netherlands \and
Max Planck Institute for Astrophysics, Karl Schwarzschild Str. 1, 85741 Garching, Germany \and
Department of Astrophysics/IMAPP, Radboud University Nijmegen, P.O. Box 9010, 6500 GL Nijmegen, The Netherlands \and
SmarterVision BV, Oostersingel 5, 9401 JX Assen \and
Th\"{u}ringer Landessternwarte, Sternwarte 5, D-07778 Tautenburg, Germany \and
Leiden Observatory, Leiden University, PO Box 9513, 2300 RA Leiden, The Netherlands \and
LPC2E - Universite d'Orleans/CNRS \and
Station de Radioastronomie de Nancay, Observatoire de Paris - CNRS/INSU, USR 704 - Univ. Orleans, OSUC , route de Souesmes, 18330 Nancay, France \and
Anton Pannekoek Institute, University of Amsterdam, Postbus 94249, 1090 GE Amsterdam, The Netherlands \and
Astronomisches Institut der Ruhr-Universit\"{a}t Bochum, Universitaetsstrasse 150, 44780 Bochum, Germany \and
Astro Space Center of the Lebedev Physical Institute, Profsoyuznaya str. 84/32, Moscow 117997, Russia \and
National Astronomical Observatory of Japan, Japan \and
Sodankyl\"{a} Geophysical Observatory, University of Oulu, T\"{a}htel\"{a}ntie 62, 99600 Sodankyl\"{a}, Finland \and
STFC Rutherford Appleton Laboratory,  Harwell Science and Innovation Campus,  Didcot  OX11 0QX, UK \and
Center for Information Technology (CIT), University of Groningen, The Netherlands \and
Centre de Recherche Astrophysique de Lyon, Observatoire de Lyon, 9 av Charles Andr\'{e}, 69561 Saint Genis Laval Cedex, France \and
Max-Planck-Institut f\"{u}r Radioastronomie, Auf dem H\"ugel 69, 53121 Bonn, Germany \and
Fakult\"{a}t f\"{u}r Physik, Universit\"{a}t Bielefeld, Postfach 100131, D-33501, Bielefeld, Germany \and
Department of Physics and Electronics, Rhodes University, PO Box 94, Grahamstown 6140, South Africa \and
SKA South Africa, 3rd Floor, The Park, Park Road, Pinelands, 7405, South Africa \and
ALMA Regional Centre Leiden Observatory, Leiden University, PO Box 9513, 2300 RA Leiden, The Netherlands 
}

\date{Received 07-10-2015 / Accepted 27-11-2015}

\abstract
{With the limited amount of in-situ particle data available for the innermost region of Jupiter's magnetosphere, Earth-based observations of the giant planets synchrotron emission remain today the sole method to scrutinize the distribution and dynamical behavior of the ultra energetic electrons magnetically trapped around the planet. Radio observations ultimately provide key information addressing the origin and control parameters of the harsh radiation environment know as of today.} 
{Here we perform the first resolved and low-frequency imaging of the synchrotron emission with LOFAR. At a frequency as low as of 127 MHz, the radiation from electrons with energies of $\sim$1--30 MeV are expected, for the first time, to be measured and mapped over a broad region of Jupiter's inner magnetosphere.}
{Measurements consist of interferometric visibilities taken during a single 10 hour rotation of the Jovian system. These visibilities were processed in a custom pipeline developed for planetary observations, combining flagging, calibration, wide-field imaging, direction-dependent calibration and specific visibility correction for planetary targets. We produced spectral image cubes of Jupiter's radiation belts at various angular, temporal and spectral resolutions from which flux densities were measured.}
{The first resolved images of Jupiter's radiation belts at 127--172 MHz are obtained, with a noise level $\sim$20--25 mJy/beam, along with total integrated flux densities. They are compared with previous observations at higher frequencies. A larger extent of the synchrotron emission source ($\geq$4 $R_\text{J}$) is measured in the LOFAR range, that is the signature -- as at higher frequencies -- of the superposition of a ``pancake" and an isotropic electron distribution. Asymmetry of east--west emission peaks is measured, as well as the longitudinal dependence of the radial distance of the belts, and the presence of a hot spot at $\lambda_\text{III}$ = 230$^\circ \pm 25^\circ$. Spectral flux density measurements are on the low side of previous (unresolved) ones, suggesting a low-frequency turnover and/or time variations of the Jovian synchrotron spectrum.
}
{LOFAR proves to be a powerful and flexible planetary imager. In the case of Jupiter, observations at 127 MHz depict the distribution of $\sim$1--30 MeV energy electrons up to $\sim$4--5 planetary radii. The similarities of the observations at 127 MHz with those at higher frequencies reinforce the conclusion that the magnetic field morphology primarily shapes the brightness distribution features of Jupiter's synchrotron emission, as well as how the radiating electrons are likely radially and latitudinally distributed inside about 2 planetary radii. Nonetheless, the detection of an emission region that extends to larger distances than at higher frequencies, combined with the overall lower flux density, yields new information on Jupiter's electron distribution, information that ultimately may shed light on the origin and mode of transport of these particles.}

\keywords{Jupiter -- radiation belts -- synchrotron emission -- radio interferometry -- LOFAR}
\titlerunning{Jupiter's synchrotron imaging with LOFAR}
\authorrunning{Girard et al.}
\maketitle
%
Version ~ : \today 

\section{Introduction}

\begin{figure}[htpb]
\resizebox{\hsize}{!}{\includegraphics{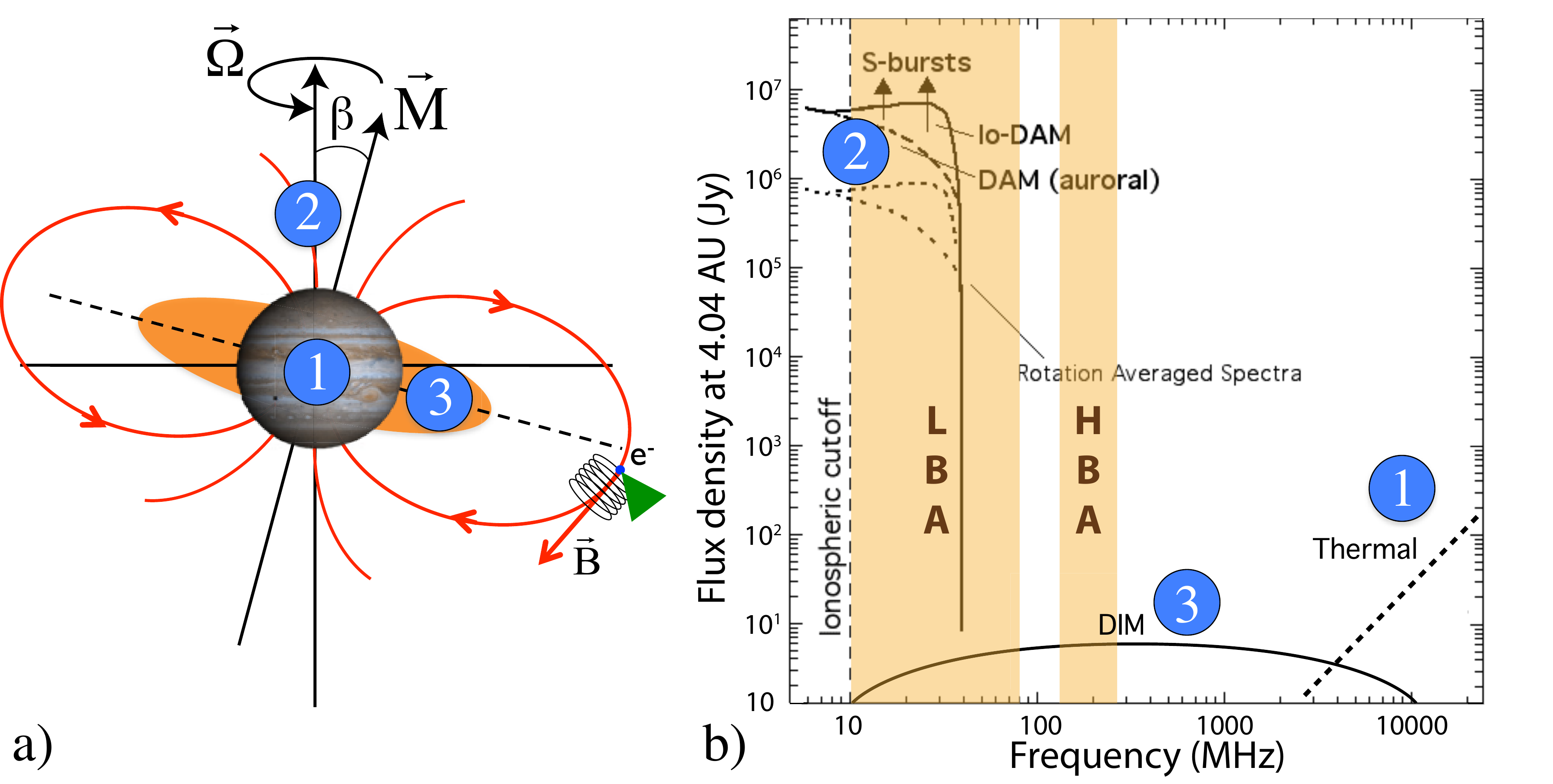}}
\caption{a) Sketch of the location of Jovian radio sources in Jupiter's inner magnetospheric field lines (in orange): (1) thermal, (2) auroral, (3) radiation belts. $\vec{\Omega}$, $\vec{B}$ and $\vec{M}$ are respectively the rotation vector, the magnetic field and the magnetic moment. The angle $\beta$ is the tilt between the rotation axis and the magnetic moment. b) Corresponding typical spectra (in Jansky normalized to 4.04 AU, adapted from \cite{Zarka_2004_FastJup}), with indication of LOFAR's low (LBA = low-band antennas) and high (HBA = high band antennas) spectral bands. DAM and DIM are the official denomination of the decameter and the decimeter emissions.}
\label{fig1}
\end{figure}

Jupiter is among the most intense radio emitters in our Solar System. It has a strong magnetic field dominated by a dipole component of moment $\sim$4.3 $R_\text{J}^3$ $\text{G}$ (1 $\text{G}=10^{-4}$ T, 1 $R_\text{J}$ = 71 492 km), much larger than that of the Earth \citep{Bagenal_2014}. This dipole is tilted by $\sim$9.6$^\circ$ relative to the rotation axis, toward a longitude of $\sim$$200^\circ$ (Fig. \ref{fig1}). The rotation of this magnetic field with a period of 9h55m29.71s defines a coordinate system of reference called ``System III'' (1965) as described in \cite{Dessler_2002}. Its interaction with the Solar wind creates a large magnetosphere, in which charged particles are accelerated to keV -- MeV energies.
Three main radio components are produced by Jupiter and its magnetosphere (Fig. \ref{fig1}): (1) the thermal emission coming from the planetary disk dominates the spectrum above $\sim$4 GHz with a brightness temperature of $\gtrsim$150 K \citep{Kloosterman_2008,Hafez_2008}; (2) auroral emission is produced below 40 MHz by electrons accelerated to keV energies in the magnetosphere at 20--50 $R_\text{J}$ from the planet, and then precipitated along magnetic field lines toward high latitudes where they produce aurorae and associated cyclotron radio emission; (3) synchrotron emission is produced between $\sim$30 MHz and $\sim$30 GHz by electrons accelerated to MeV energies and trapped in the so-called radiation belts of the inner magnetosphere, within a few radii of the surface, mostly at low latitudes. In the present paper, we are interested in this synchrotron emission.

Since its discovery in the mid-fifties, the synchrotron emission has been imaged between 330 MHz and 22 GHz using various instruments (VLA, WSRT, ATCA, \dots), and a few unresolved measurements have also been performed down to 74 MHz (VLA, CLFST) \cite[see][]{dePater_2003c}. Ground-based synchrotron measurements provide valuable information about the angular and frequency distribution of high-energy electrons trapped in Jupiter's inner radiation belt (<6 $R_\text{J}$). Relying on the well understood physics of synchrotron emission, they are used to test physical models of the radiation belts, incorporating various physical processes such as radial diffusion of the electrons, interaction with the magnetospheric plasma, satellites, rings and plasma waves, and synchrotron losses \citep[see e.g.,][]{dePater_1981,dePater_1997,SantosCosta_2001b,Bolton_2004,dePater_2004,SantosCosta_2008}.

Energetic electrons are in fast gyration around Jupiter's magnetic field lines. This gyration motion is associated to an invariant that is the magnetic moment of the electron $E_\bot/B$, that causes bouncing of the electrons between magnetic mirror points where the parallel velocity reverses and the pitch angle is $\sim$90$^\circ$. The synchrotron emission taps the perpendicular energy of the electrons and is beamed in the direction of electron motion. As a consequence, the bulk of emission comes from electrons having a velocity near-perpendicular to magnetic field lines. Because the Earth always lies within $\sim$13$^\circ$ of the Jovian magnetic equator, the field lines are themselves near-perpendicular to the line of sight (Fig. \ref{fig1}). An accumulation of such particles exists around the magnetic equator (for the electrons with a large equatorial pitch angle, trapped between magnetic mirror points at low latitudes) and at high magnetic latitudes (where the mirror points of energetic electrons with small pitch angles lie; due to their small parallel velocity there, electrons reside a long time near these mirror points, leading to enhanced synchrotron emission). Since the emitted power is proportional to $E^2 \times B^2$, the peak frequency proportional to $E^2 \times B$, with $E$ the electron's energy and $B$ the magnetic field strength at the source, synchrotron spectra as well as images at different frequencies allow us to probe the distribution of electrons at various energies in Jupiter's inner magnetosphere. Lower radio frequencies are associated with lower energy electrons (typically several MeV) in a strong B field and higher energy electrons at greater distances from the planet (i.e., in a weaker magnetic field). Hence, it is difficult to disentangle the energy distribution of the electron in observations without any spatial resolution, since this information is entangled with information about the pitch angle distribution and the line-of-sight integration through a complicated magnetic field configuration. High resolution imaging is thus crucial to derive sound constraints. No resolved image has been obtained yet below 330 MHz \citep{dePater_2004}. It is in particular interesting to map Jupiter at frequencies in the 70--300 MHz range since the disk-integrated spectral measurements are suggestive of a turnover in the spectrum at these lower frequencies \citep{dePater_2003b}.
Moreover, at LOFAR frequencies, resolved imaging is valuable as it enables ``scanning'' for the first time the 1--30 MeV electron population through their contribution to the synchrotron emission located further away from the planet. In the equatorial plane, assuming a dipolar magnetic field in the region from 1 $R_\text{J}$ to 4 $R_\text{J}$, the detectable synchrotron emission at 1.4 GHz is associated to electrons with energy ranging from 7.9 MeV up to 67 MeV. With a rule of thumb, at 127 MHz, in the same region, we can probe electrons populations from to 2 MeV up to 20 MeV. Therefore, the study of the resolved synchrotron emission with LOFAR at low frequencies and in distant regions of the belts contribute to constraint the electron populations originating from the middle magnetosphere and undergoing inward diffusion and acceleration processes.
In this paper, we present the first resolved images of Jupiter's synchrotron emission obtained (with LOFAR) at a frequency as low as 127 MHz, as well as disk-integrated spectral measurements, and we derive preliminary constraints on the morphology and variability of the emission. Observations and the custom pipeline that we developed for analyzing LOFAR planetary observations are described in Section 2. The resulting images and spectrum are presented in Section 3, and quantitatively analyzed in Section 4. Section 5 discusses these first low-frequency observations and perspectives for further studies.

\section{Observations and planetary imaging pipeline}

\subsection{Observational requirements for planetary imaging}
Planetary imaging requires a special observing strategy and calibration as compared to other radio observations. For Jupiter, two main effects have to be taken into account: i) the proper motion of Jupiter on the sky background, ii) the intrinsic motion of the radiation belts in the reference frame of the planet.

First, as we observe from the ground, Earth's (and Jupiter's) orbital motion induces an additional apparent motion of planetary targets with respect the rest of the sky, the apparent motion of which is due to Earth's rotation. This cause the planetary source to travel over the course of the year between radio sources with the consequence of impacting the calibration of long integrated observations. This motion is relatively fast for Jupiter, causing a shift of 3.16 arcminutes -- i.e., nearly 4 times its diameter -- during one 10h planetary rotation, relative to the ``fixed'' RA/DEC sources (e.g., NVSS source J020457+114145). Although this is large compared to our $\sim$7'' synthesized beam width, it is much smaller than the $\sim$5$^\circ$ primary beam of the telescope).

\begin{figure*}[h!]
\resizebox{\hsize}{!}{\includegraphics{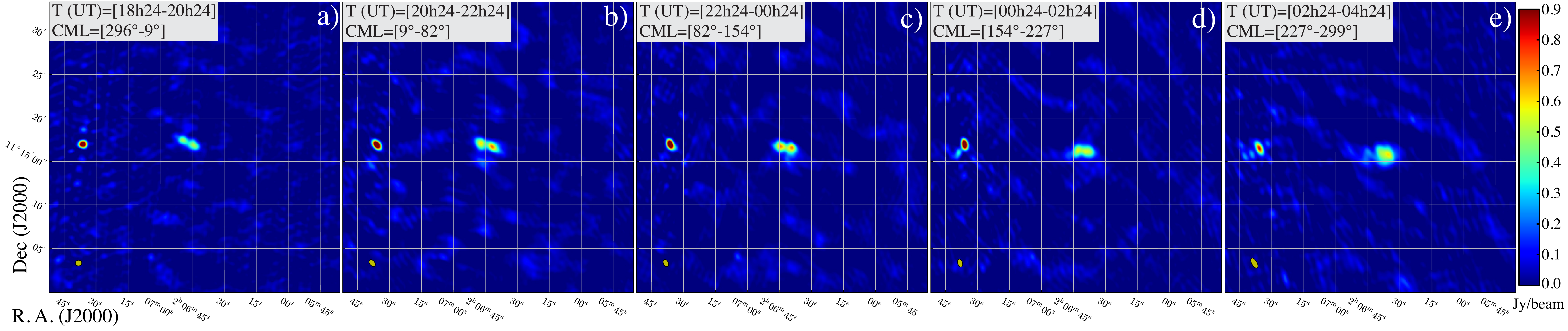}}
\caption{Calibrated images of the Jupiter in a 35'$\times$35' field, integrated over 20 sub-bands of 195 kHz within the range 166--172 MHz, for five consecutive 2 hours intervals (indicated on each panel). 
The selected subset of baselines provides an angular resolution of 35". Pixel size is 5". 
All panels are centered around the target phase direction (in RA/DEC). The belts are resolved near the center of the image and the beam is displayed in the bottom-left corner in yellow. The point source, east of Jupiter, is J020457+114145. Observing time and CML (CML = central meridian longitude in System III) are indicated on each panel. The motion of Jupiter from image to image is clearly visible. The rocking of the main axis of the image of Jupiter is also discernable. The last image is more distorted, due to lower quality data in the last time window.}
\label{fig2}
\end{figure*}

\begin{figure}[!ht]
\resizebox{\hsize}{!}{\includegraphics{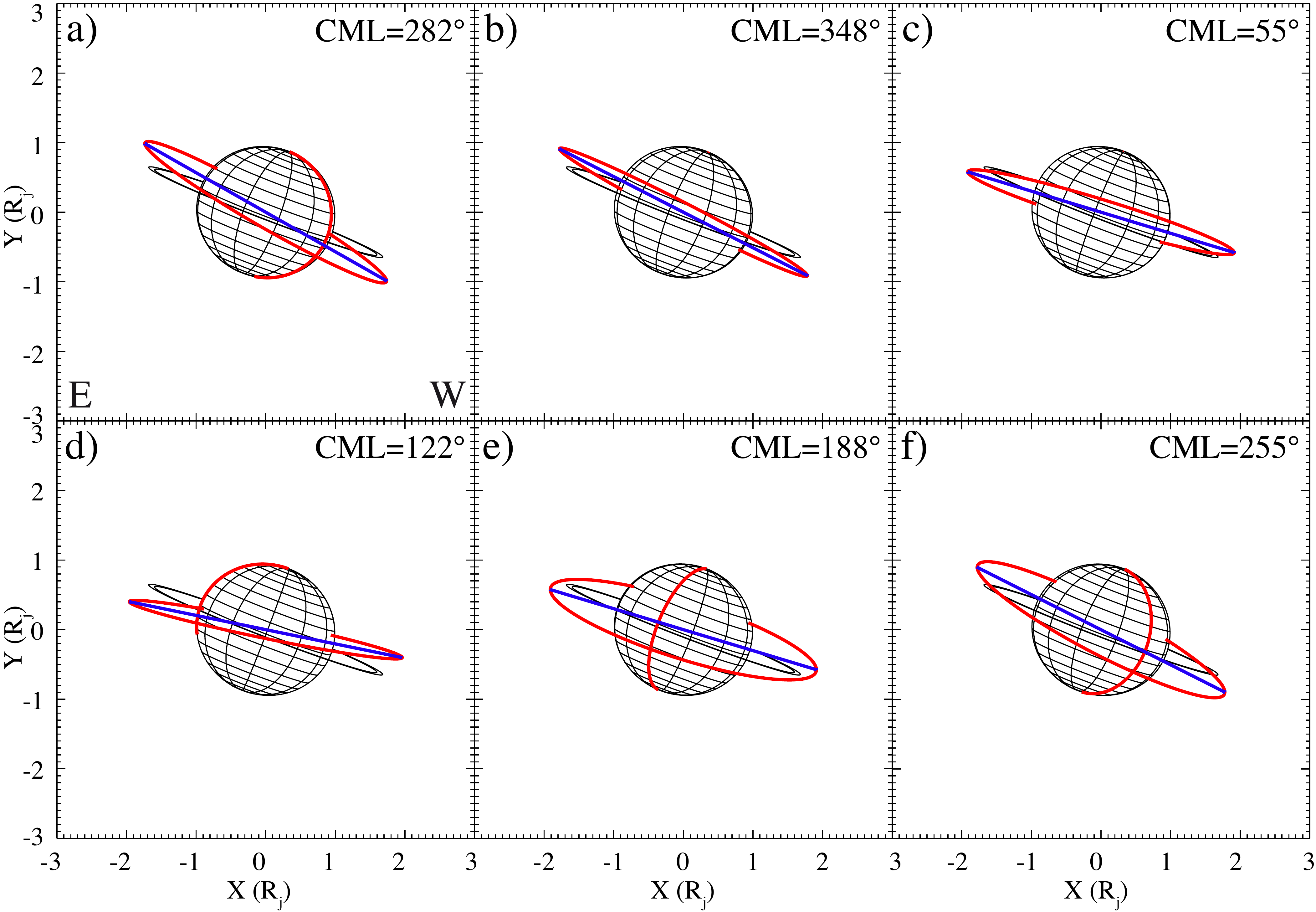}}
\caption{Wobbling of the Jovian magnetic equator (red circle) on the plane of the sky during one planetary rotation. The rings (in black) lie in the rotational equator. The blue line represents the main axis of the projection on the sky of the magnetic equator, which should also be the main axis of the radiation belts image as a function of time or longitude. Observer's longitude (CML) is indicated on each panel as well as the reference meridian (in red).}
\label{fig3}
\end{figure}

Second, Jupiter's radiation belts are fixed -- at zero order -- relative to the Jovian magnetic field. But as Jupiter's magnetic dipole axis is tilted by 9.6$^\circ$ with respect to its rotation axis, the former precesses around the latter with Jupiter's rotation (see Fig. \ref{fig1}). As a consequence, the magnetic equator and the whole image of the radiation belts wobbles or rocks on the plane of the sky at the System III period (Fig. \ref{fig3}). This rocking is discernable between panel a) and d) of Fig. \ref{fig2}, where the main axis of the image of Jupiter has changed in orientation.

As a third and minor effect, the varying distance between the observer and Jupiter has to be taken into consideration. Therefore, all measured flux densities must be scaled to the common reference distance of 4.04 AU to enable comparison between epochs.

Jupiter's synchrotron emission is a few jansky (1 Jy = $10^{-26}$ Wm$^{-2}$Hz$^{-1}$) radio source that is resolved by LOFAR; therefore a long time-integration is necessary to obtain an image with a good signal-to-noise ratio (SNR, defined as the ratio of the peak flux to the background r.m.s. noise). If no precaution is taken while producing long time and frequency integrated images, the displacement motion of Jupiter and the rocking motion of its radiation belts will lead to a large smearing of their image (in addition to time and frequency smearing effects which also slightly distorts the shape of the sources located at the edge of the beam). The former motion must be compensated in the Fourier domain via a time-dependent translation of the phase center including antenna delay correction, and the latter by a rotation of the visibility reference frame, i.e., the ($u$,$v$) axes, prior to imaging. However, these corrections should only be applied to Jupiter visibility data, otherwise they will cause a systematic smearing of other fixed coordinate radio sources in the field, increasing the difficulty of imaging sources that are no longer point sources. 

Therefore, to enable posterior correction of these effects, an observation should be carried toward an arbitrary pointing direction with a constant RA/DEC coordinate.
As illustrated in Fig. \ref{fig2}, the target beam was pointed to Jupiter's mean RA/DEC position during the full observing period. We can see the motion and rocking effects in the preliminary LOFAR images derived for five consecutive 2 hours intervals.

\subsection{Observational setup}
The observations analyzed here were recorded during the commissioning period of LOFAR. They consist of visibilities recorded within a 10 hour window from 18:24 UT on 2011/11/10 to 04:24 UT on 2011/11/11. At the observation epoch, Jupiter was at 3.99 AU from Earth and subtended an angle of $\sim$49" in the sky. The Sun--Earth--Jupiter angle was 165$^\circ$ and the Earth was at a Jovigraphic latitude ($D_\text{E}$) of +3.29$^\circ$ (\texttt{www.imcce.fr}). 
Observations were carried out using 49 High Band Antenna (HBA) fields (2 per station for 20 Core Stations + 1 per station for 9 Dutch Remote Stations \citep[see][]{vanHaarlem_2013}. Two station beams were synthesized by phasing the antennas at station level: a ``target'' beam centered on Jupiter, and a ``calibration'' beam centered on the radio source 4C15.05 (for phase calibration) four degrees away from Jupiter. The approximate half power beam width (HPBW) of the beams is $\sim$5$^\circ$ at 150 MHz. The same $\sim$23 MHz of total bandwidth were recorded from each beam, in the form of 121 sub-bands of 195 kHz, in twelve groups of ten contiguous subbands (each group is therefore 2 MHz wide), regularly distributed between 127 and 172 MHz (hence a spectral coverage of 50\%). The raw data consist of complex visibilities produced at $\sim$1 s time resolution and in 3 kHz-wide frequency channels for all available baselines. Baseline lengths were distributed between $\sim$15 $\lambda$ and $\sim$30 k$\lambda$ (with $\lambda$ the wavelength). The ($u$,$v$) radial density peaks at $\sim$500 $\lambda$ (corresponding to Core Station baselines) and is then approximately flat up to 30 k$\lambda$, providing a maximum theoretical angular resolution of $\sim$6.5". The two co-polarization ($XX$, $YY$) and two cross-polarization ($XY$, $YX$) terms were recorded, but only total intensity $I$ measurements were reliable at this early stage of LOFAR exploitation, thus we limit our present analysis to those measurements and we did not exploit the $Q$, $U$ and $V$ data. 

\subsection{Flagging and direction-independent calibration}
A classical data pre-processing was applied \citep{vanHaarlem_2013}: flagging of radio frequency interference (RFI) using the AOflagger \citep{Offringa_2012} followed by time integration on 3 s steps and by frequency integration on 195 kHz (1 LOFAR sub-band) by the LOFAR NDPPP pipeline \citep{Pizzo_2015}, calibration using the phase calibrator field using BBS \citep{BBS}, then derivation of complex gain solutions for all antennas in 9 s bins (i.e., 1 gain solution every 3 time bins). Gain amplitudes and phases were then visually inspected and bad data were flagged. The gain solutions were significantly more noisy during approximately the first and last hour of the observation (due to the low elevation of the source and probably the ionosphere turbulence state).
Strong radio sources such as the ``A''-team (e.g., Cas A, Cyg A, Vir A, Tau A,\dots) can contaminate LOFAR data if they are present in the station side lobes. As the HBA band is less affected than the LBA by the A-team, and as visual inspection of visibilities did not reveal the contribution of any A-team source in the data, we did not apply any specific treatment to these strong radio sources.

\subsection{Direction-dependent calibration, background subtraction and proper motion correction}
To alleviate the spatial smearing caused by the planetary corrections in the visibility plane, we need to detect and subtract all other radio sources in the data to improve the dynamic range of the image of the target. Because the field of view (FoV) of the LOFAR stations is large ($\sim$5$^{\circ}$ HPBW at $\sim$150 MHz), wide-field imaging within the full FoV is required.

Thus, the planetary imaging pipeline that we developed includes the following steps: (i) make a wide-field image of the target field from the calibrated visibilities; (ii) detect in the image the sources other than Jupiter above a given threshold and identify them using a radio source catalog;
(iii) subtract these sources with direction-dependent (DD) calibration solutions ; (iv) apply the above motion corrections to the peeled visibilities and ($u$,$v$,$w$) coordinates; (v) build final Jupiter images integrated over selected intervals of time and frequency.

\begin{figure}[ht]
\centering
\resizebox{0.75\hsize}{!}{\includegraphics{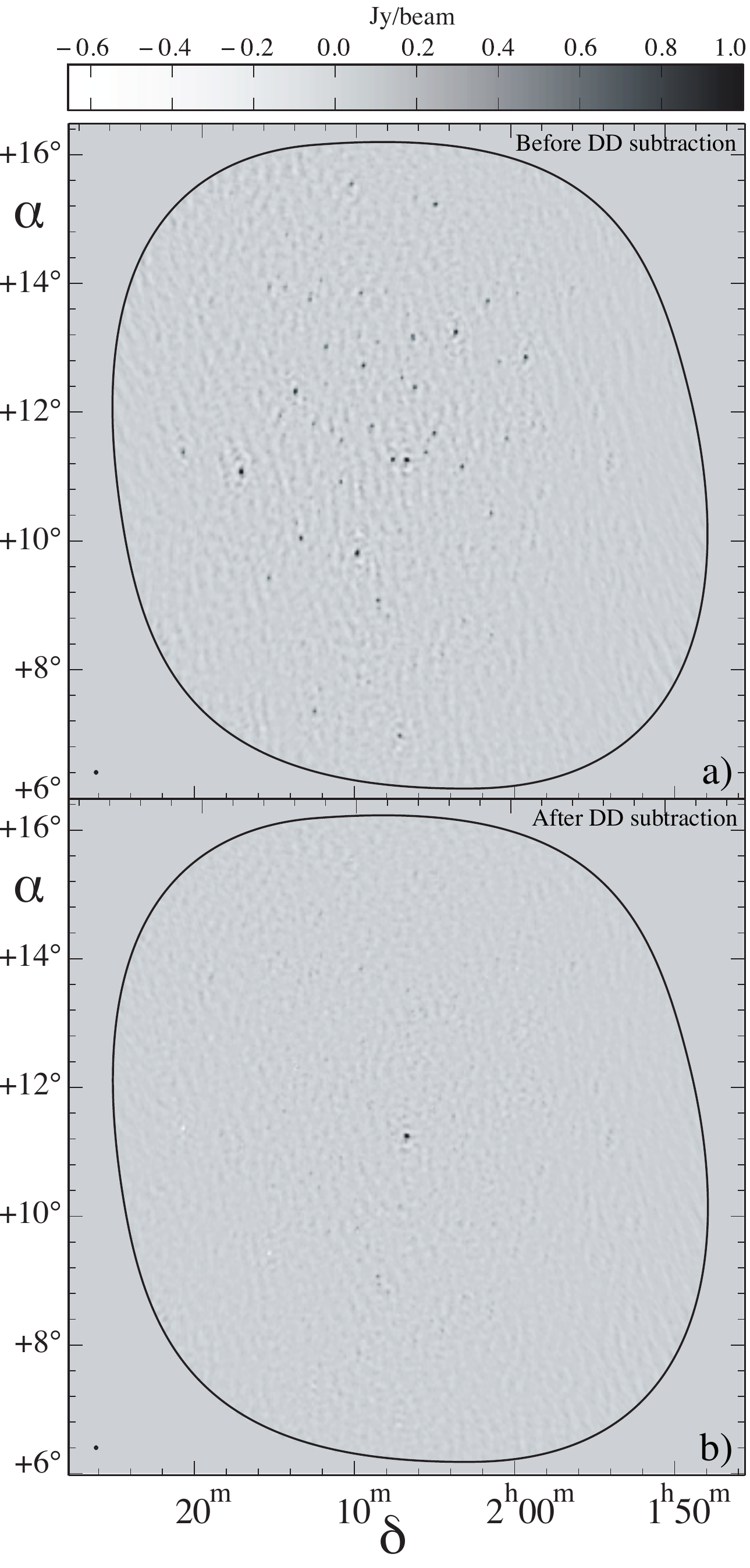}}
\caption{Wide-field ($8^\circ \times 8^\circ$) image of the target field before a) and after b) DD subtraction. The central frequency is 143 MHz and the bandwidth is 1.9 MHz (SB 40--49). The image is 2100$\times$2100 pixels of 18", baseline length was restricted to $\leq$10 k$\lambda$ to ensure a good sampling of the PSF. A natural weighting was used and deconvolution with AWimager (implementing A-projection and W-projection) with 10 000 iterations and a CLEAN gain of 0.1. The angular resolution is $\sim$2.5'. The r.m.s. and the SNR are respectively 37 mJy/Beam and 71 in the non-peeled image (a) and  $\sim$27 mJy/Beam and 98 in the peeled image (b). Sixty background sources have been automatically identified and subtracted in panel a) to obtain the image of panel b). Jupiter lies at the center of the field. The beam size and shape are displayed at the bottom left of each panel.}
\label{fig4}
\end{figure}

For building the wide-field image (i) we used the AWImager \citep{Tasse_2013} that does beam correction (A-projection) and wide-field imaging corrections \citep[W-projection, ][]{Cornwell_2008}. Automatic source detection (ii) was performed using the Duchamp source finder \citep{Whiting_2012} and a sky model creator \texttt{buildsky} \citep[][and references therein]{Yatawatta_2013}.
Most of the detected sources could be associated with the GSM \citep[Global Sky Model,][]{Pizzo_2015} that contains radio sources from the VLSS, NVSS and WENSS surveys. The GSM provides a realistic model of the sky with reliable flux densities and spectral indices. However at LOFAR wavelengths, the spectral index of some radio sources decreases, which introduces a systematic bias when their flux densities are extrapolated from high frequencies. Moreover, we assume here that the sources in the catalog are not variable in time. Thus at step (iii) we chose to subtract the sources with their observed flux density in each 2-MHz-bandwidth image, using the experimental DD source subtraction algorithm (\texttt{Cohjones} developed by \citet{Tasse_2014,Tasse_2015}) accounting for the beam variations. Steps (i) to (iii) are illustrated in Fig. \ref{fig4} that displays a wide-field ($8^\circ \times 8^\circ$) image of the target field before and after DD subtraction. A total of 60 sources split in 8 clusters (e.g., 8 directions) down to 0.2 Jy have been automatically identified and subtracted from the visibilities in panel a) to obtain the image of panel b). In panel b), Jupiter (unresolved) is the dominant source in the visibility data. Source residuals are visible at the location of each removed source but their contribution to the noise (i.e. the calibration and deconvolution noise) has been strongly reduced and they are relatively far from the region of interest. Source subtraction allowed us to reduce the r.m.s. noise by $\sim$ 30\% in each frequency band. 
Step (iv) is detailed in the appendix and the combined motion of the radiation belts was corrected for every 5 minute window of the observation \citep[details can be found in][]{Girard_2012,Girard_2012b}.

Step (v) consist of deconvolution and image cube creation. At step (i), we used AWImager \cite{Tasse_2013} to perform wide field imaging of all sources with beam corrections and W-projection. After steps (iii) and (iv), the visibilities are mainly dominated by the synchrotron emission from the radiation belts in a small region near the center of the field. Therefore, we used the Cotton-Schwab CLEAN algorithm implemented in CASA \citep{NRAOcasa} to produce final images of the radiation itself.

\subsection{Image and spectrum processing of Jupiter's radiation belts}

We have built a $12 \times 5$ image cube (one image per 2 MHz band and per 2 hours of integration), 5 frequency-averaged images (one image per 2 hours, integrated over the 23 MHz of bandwidth between 127 and 172 MHz), and 12 rotation-averaged images (one image per 2 MHz band, integrated over $\sim$7 hours -- from $\sim$19:00 to $\sim$02:00 UT). 

\begin{figure*}[!t]
\centering
\resizebox{\hsize}{!}{\includegraphics{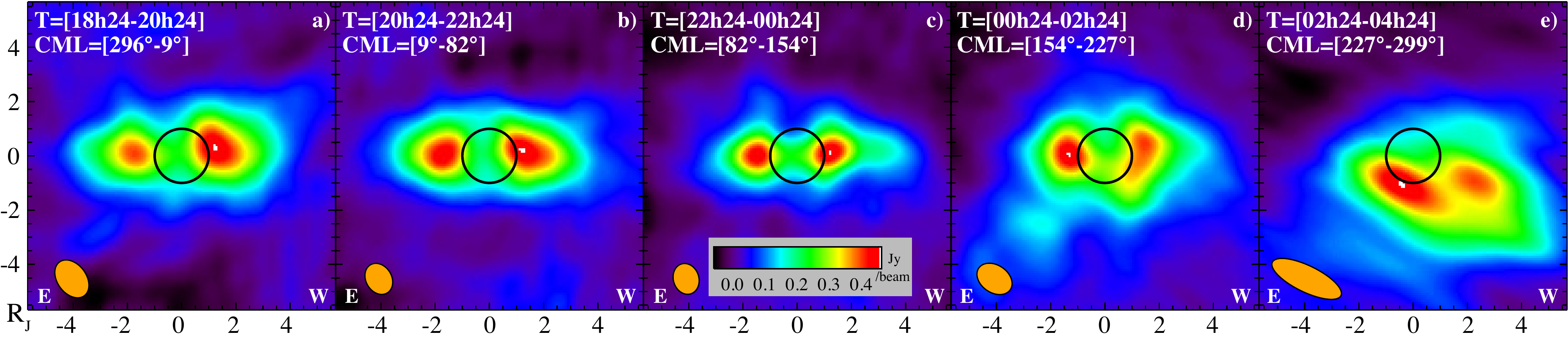}}
\caption{Zoomed images of Jupiter's radiation belts obtained after motion corrections in the ($u$,$v$) plane and DD subtraction, in a $\sim$10'$\times$10' field, integrated over the whole bandwidth for the same time intervals as Fig. \ref{fig2}. The spatial scale is given in Jovian radii at Jupiter (1 $R_\text{J}$=71492 km, corresponding to $\sim$49" in the sky at observation epoch). Imaging was performed with baselines $\leq$15 k$\lambda$, giving a theoretical angular resolution of 14" and an effective angular resolution ranging from 20" to 78" over the 10 hours. Pixel size is 2"$\times$2". In the five successive images a) to e), the residual noise level is 14.9, 10.5, 12.3, 15.9 and 21.2 mJy/beam, and the peak SNR ratio is respectively 31.0, 33.6, 34.1, 26.0 and 17.6. The last image is more distorted because of the low source elevation during that observing interval.}
\label{fig5}
\end{figure*}

\begin{figure*}[!t]

  \centering
\resizebox{\hsize}{!}{\includegraphics{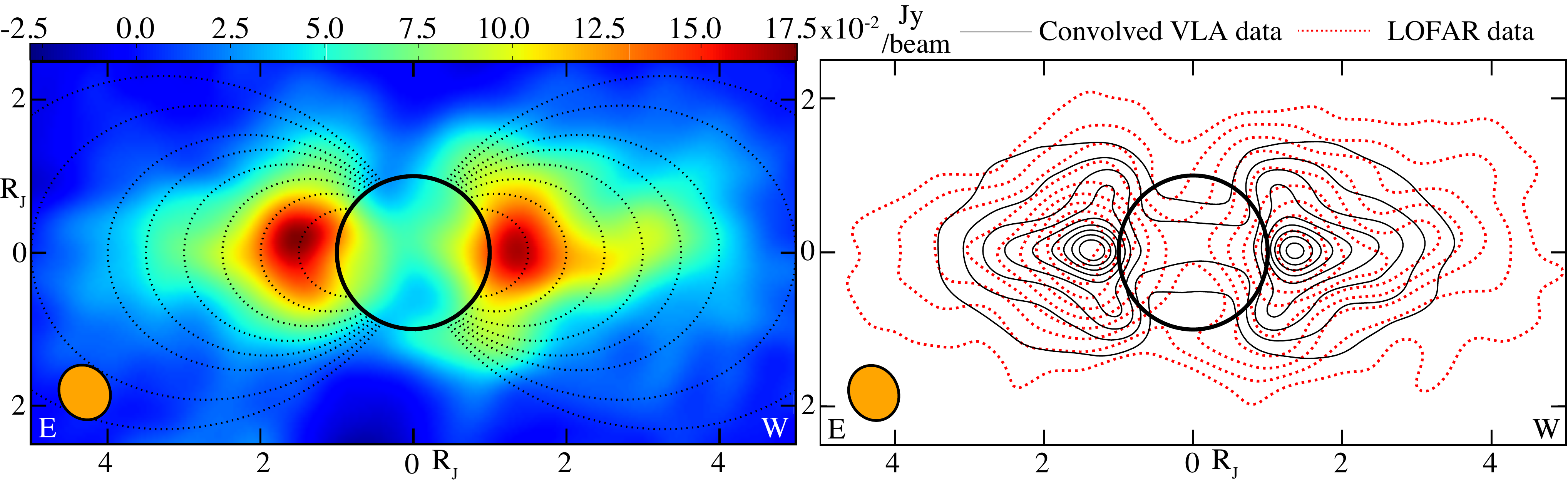}}

\caption{Best LOFAR high resolution image to date of Jupiter's radiation belts, integrated over the entire 23 MHz band of observation distributed from 127 and 172 MHz, and a 7 hour interval from $\sim$19:00 to $\sim$02:00 UT. The same image is displayed in both panels with color scale (left) and contours (right). The frequency-averaged clean beam size and shape ($\sim$18"$\times$16") is displayed at the bottom left of each panel. Pixel size is 1"$\times$1". R.m.s. noise is 4.7 mJy/beam and the SNR (maximum peak flux divided by standard deviation) is 37. The SNR is approximately 14 at the 30 \% flux level (corresponding to the extremity of the emission). Dipolar field lines with apex at 1.5, 2, 2.5, 3, 3.5, 4 \& 5  $R_\text{J}$ are superimposed on the left panel. Contours superimposed on the right panel are derived from a rotation-averaged VLA image \citep[obtained from][in C band]{SantosCosta_2009} which has been convolved down to match the LOFAR observation angular resolution. Each set of contours represents relative intensity levels by steps of 10 \% of the maximum radiation peak: in the convolved VLA image (black line) and in the LOFAR image (red dotted line).}
  \label{fig6}
\end{figure*}

The five frequency-averaged images are displayed in Fig. \ref{fig5}, centered on the position of Jupiter. We can see that the shift and the rocking of the radiation belts have indeed been corrected. We also note that the detailed shape of the radiation belts varies from panel to panel, which we attribute to the limited SNR of each image. 
The last image is very distorted, due to the noisy character of the last portion of the data and the poor ($u$,$v$) coverage due to the low elevation of the source at the end of the observation ($\sim$10$^{\circ}$). A more detailed analysis of intermediate images shows that the interval with highest quality data is the 7 hour interval from $\sim$19:00 to $\sim$02:00 UT, that we used for building the 12 rotation-averaged images (not displayed). Finally, from these 7 hours and the entire 23 MHz bandwidth of observation, we built the time-and-frequency-averaged image of Fig. \ref{fig6}, which is the first resolved image of Jupiter obtained in the 127--172 MHz band. The residual noise in this image is 4.7 mJy/beam, giving a peak SNR of 37. At the extremity of the extended emission (around the 30 \% of the peak flux), the ``local'' SNR ratio is $\sim$14.

\begin{figure}[!ht]
\centering
\resizebox{\hsize}{!}{\includegraphics{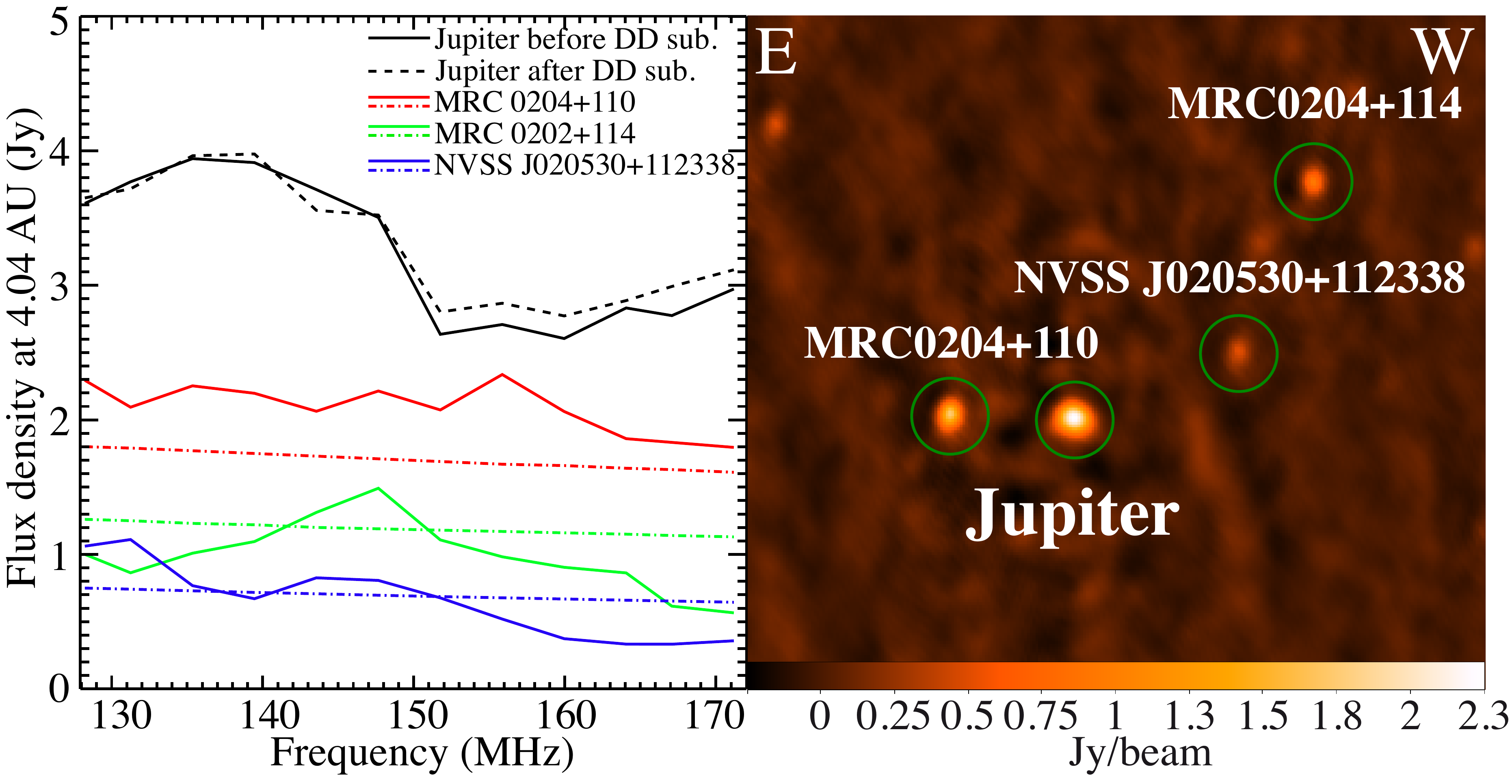}}
\caption{a) Wide-field unresolved image of Jupiter and its vicinity (zoom in a region of $\sim$1.2$^o\times$1.2$^o$) before source subtraction. The 3 bright nearby radio sources have the following flux density S at 73.8 MHz and spectral index $\alpha$ (following the convention $S_\nu \propto \nu^\alpha$): S=2.24 $\pm$ 0.23 Jy and $\alpha$ =$-$1.0 for MRC 0204+110 ; S=1.00 $\pm$ 0.12 Jy and $\alpha$ = $-$0.9 for NVSS J020530+112338 ; S=1.94 $\pm$ 0.17 Jy and $\alpha$ = $-$0.9 for MRC 0202+114 (see NED database \texttt{ned.ipac.caltech.edu}). b) Measured spectra of these 3 radio sources at 12 frequencies between 127 and 172 MHz (solid color lines) compared with NED predictions (dash-dot color lines). The black lines show the total integrated flux density measured on the unresolved image of Jupiter before and after DD subtraction of surrounding radio sources at the same 12 frequencies. The difference between the two black lines is marginal.}
\label{fig7}
\end{figure}

\begin{figure}[!ht]
\resizebox{\hsize}{!}{\includegraphics{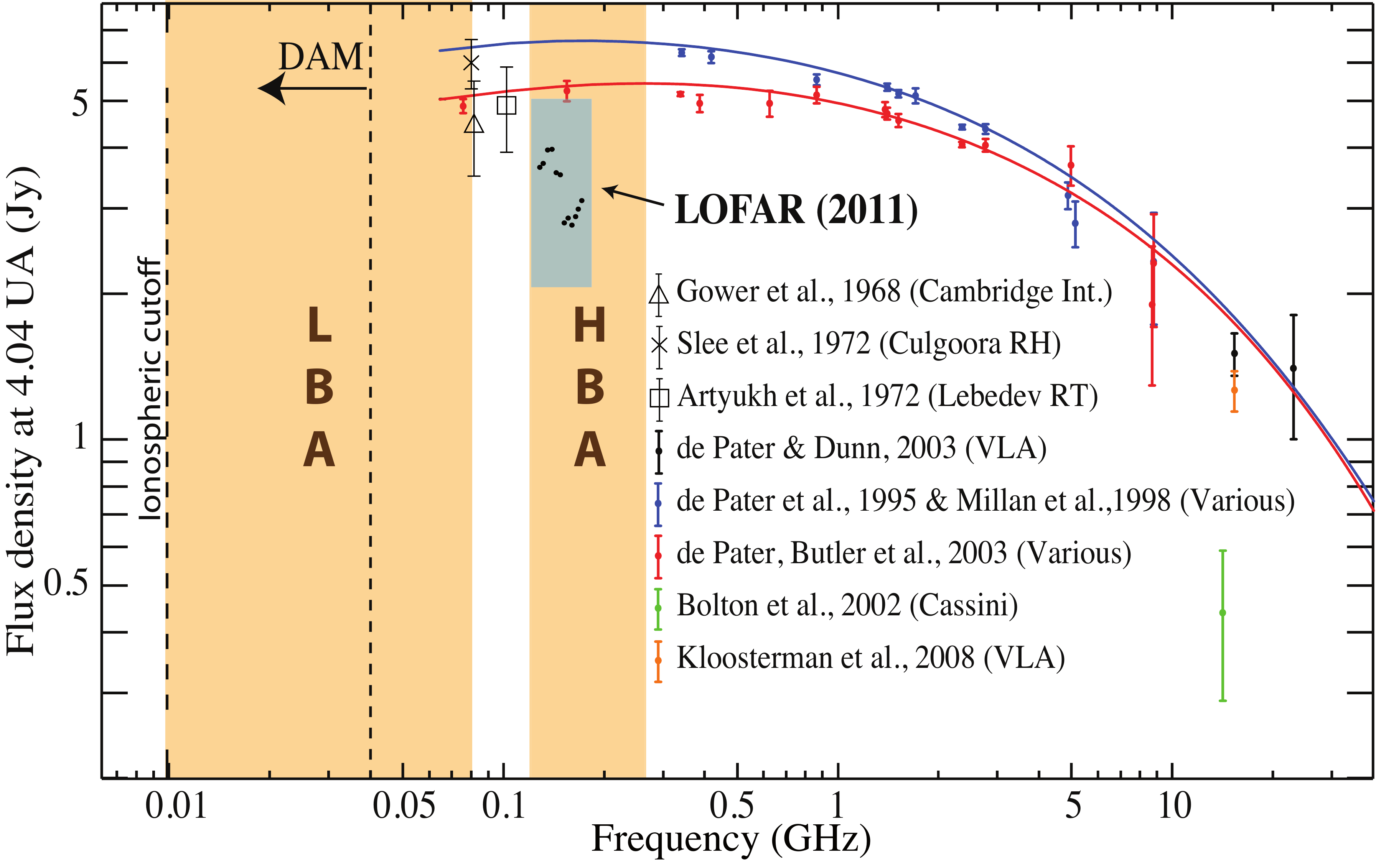}}
\caption{Measurements of Jupiter's synchrotron spectrum at meter to centimeter wavelengths, scaled to the distance of 4.04 AU. The majority of the measurements was obtained with the VLA between 1991 and 2004 \citep{dePater_1995,Millan_1998,dePater_2003a,dePater_2003b,dePater_2003c,Kloosterman_2008}, some of which were after the impact of Comet Shoemaker-Levy 9 in 1994. The blue and red curves were fitted by \citet{dePater_2003c} to the series of measurements taken respectively in 1994 and 1998. LOFAR measurements of the present study are the black dots in the HBA range, and their uncertainty is figured by the grey box. Previous measurements below $\sim$300 MHz are unresolved. Decameter emission (DAM -- not shown) dominates the spectral range below 40 MHz (see Fig. \ref{fig1}b).}
\label{fig8}
\end{figure}

In order to calibrate the flux density in the images, and to derive total integrated flux densities over the entire radiation belts that can be compared to previous measurements, we have performed source-integrated flux measurements similarly on Jupiter and on 3 nearby sources (Fig. \ref{fig7}a), before the DD subtraction step (iii), in each of the twelve 2 MHz bands. We have compared the measured total flux at each frequency with the spectra deduced from the catalogued fluxes and spectral indices for the 3 nearby sources. Measured values lie within 30 \% of values deduced from catalogues for all 3 sources surrounding Jupiter (Fig.~\ref{fig7}b), which is compatible with the uncertainty of the absolute flux densities of the GSM source at LOFAR wavelengths. We take this 30 \% value as a good measure of the maximum relative uncertainty on our total flux density measurements on Jupiter. 
No specific fitting of the beaming curve (as a function of CML) has been done on the LOFAR data (as in \cite{dePater_1989}) to measure the $A_0$ parameter corresponding to the total mean flux density over a rotation. We assume that the total flux density measured after frequency integration (next section) is representative of the mean value and close to enough to $A_0$ considering the overall uncertainty of the flux density.
The corresponding total integrated flux densities and their uncertainties are displayed in Fig. \ref{fig8}, together with previously published measurements.

\section{Analysis of the integrated images, spectra, flux variability and beaming}

\begin{figure}[!t]
\resizebox{\hsize}{!}{\includegraphics{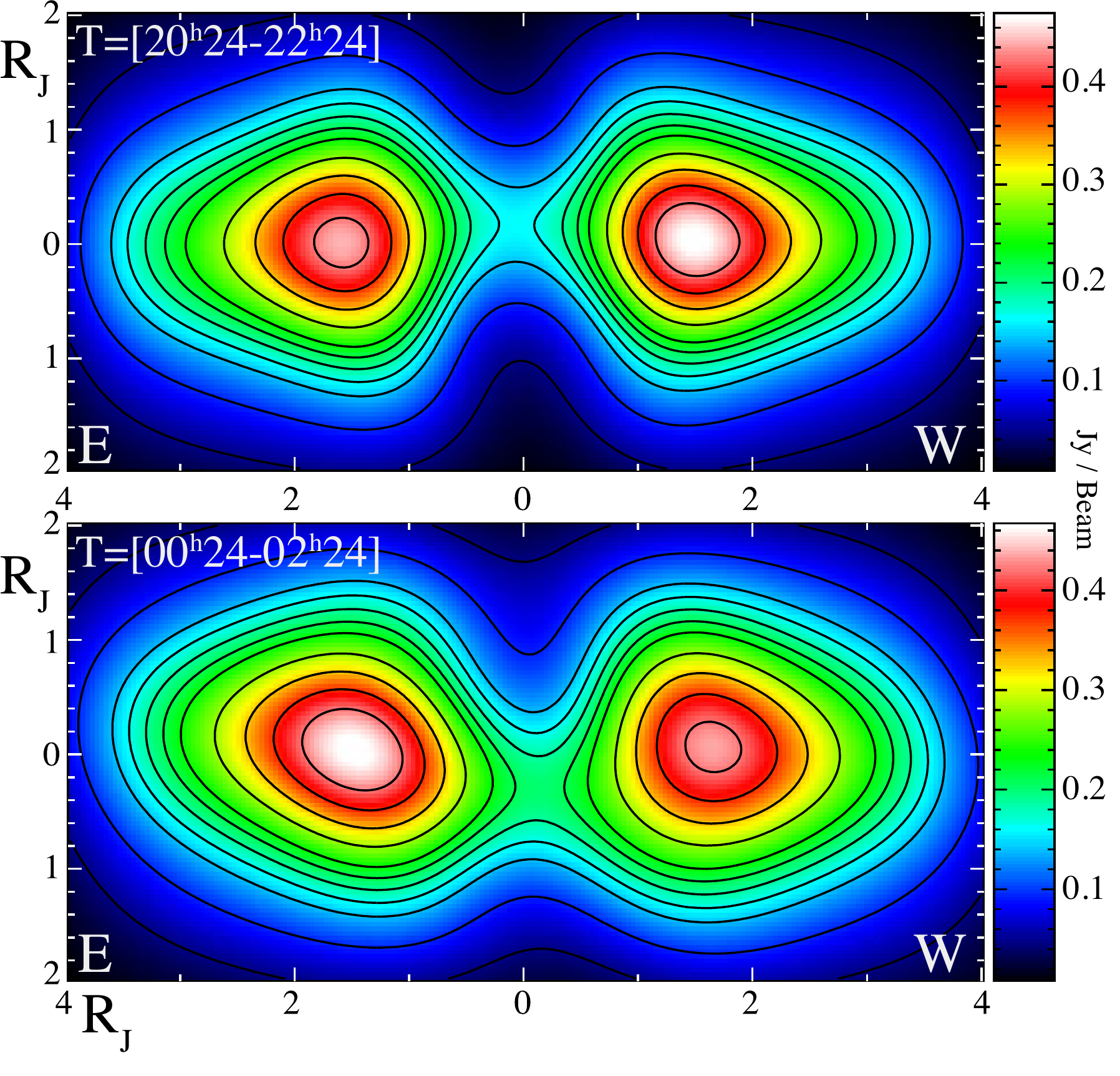}}
\caption{Simulated synchrotron images derived from Salammb\^o-3D (see text) of the emission averaged over the full band at two periods: i) [20h24--22h24] UT (CML=[9$^\circ$--82$^\circ$]), min/max=[$8\times10^{-3}$,0.475] Jy/beam and ii) [00h24--02h24] UT (CML=[154$^\circ$--227$^\circ$]), min/max=[$8\times10^{-3}$,0.467] Jy/beam. The synchrotron maps were computed with the same observation parameters as in Fig. \ref{fig5}b) and d). Contours highlight the brightness by steps of 10\% of the maximum.}
\label{fig9b}
\end{figure}

We have carried out a first analysis of the LOFAR images and spectrum. In Fig. \ref{fig5}, after planetary motion and wobble correction, we note that the brightness maximum peak is first located on the west side of the planet and is located in the east side approximately half a rotation later. This effect is relatively well known and associated to the ``beaming'' curve highlighting the variation of the peak maxima with CML over a rotation and depending of the observing geometry (controlled by the observer latitude $D_\text{E}$, \citep{Dulk_1999b}. To backup this assertion, modeling of the electron population is required as well as a synchrotron model. We reproduced the situations of panel b) and d) of Fig. \ref{fig5} with simulated synchrotron images (Fig. \ref{fig9b}) derived from Salammb\^o-3D particle code coupled with a synchrotron imaging model taking into account LOFAR observation parameters (time/CML coverage, frequency band and angular resolution). Salammb\^o-3D was originally developed for Earth radiation belts computation, but was later adapted to Jupiter's belt system and used to study the dynamics of inner belts \citep{Bourdarie1996,SantosCosta_2001,SantosCosta_2001b,Sicard_2004,Sicard_2004b}. 
At present, the code uses the [O6 + Khurana] coupled magnetic model \citep{Khurana_1997} and models the dynamics of electrons from 0.025 to 712 MeV, each contributing to the synchrotron emission at different frequencies and in different regions of the inner belts. While being refined for Earth and Jupiter, it was also adapted to other magnetic bodies such as Saturn \citep{Lorenzato_2012}.
The output of the simulation (assuming an infinite angular resolution) was convolved by the median beam over the whole frequency range in the two CML ranges. These simulations show that the maximum brightness peak effectively changes sides from west to east over few hours, consistent with the LOFAR observation in Fig. \ref{fig5}. The radial position of the brightness peaks and the extent of the belts are also consistent with the observation. 
Such preliminary comparisons suggest that a further quantitative investigation at all CMLs with detailed fitting of the physical model of the radiation belts (and of the electron populations) can lead to an accurate understanding of their morphology at low frequencies, an area that is fairly unexplored in a resolved regime of time, frequency and angular resolution. Exploitation of new wide-band data (using LOFAR \& WSRT) and a complete modeling using this kind of particle code is currently ongoing and will be subject to a future publication.

 In Fig. \ref{fig6} (right), the mean synchrotron emission appears to extend above the noise level up to a distance of $\geq$4 $R_\text{J}$ of Jupiter's center, farther out than images at higher frequency (represented as contours derived from C-band VLA data taken in 1997 by \citet{SantosCosta_2009} and convolved down to the LOFAR angular resolution). Especially, at a distance of $\sim$3.5 $R_\text{J}$,  the brightness is 10\% of the peak flux at high frequency whereas at low frequencies, still 30\%--40\% of the brightness is present at the same location, suggesting a larger extent of the radiation belts at low frequencies.
This concurs with the samples of in-situ particle data collected by Pioneers 10/11 and the Galileo probe/orbiter in the early 1970s and late 1990s and early 2000s, which have showed dense population of electrons with energies of $\sim$1--30 MeV in Jupiter's inner magnetosphere.
 This is compatible with VLA observations of \cite{SantosCosta_2014} where the radiation zone of Jupiter at P band is observed to be slightly more extended than at L and C bands (quiet state or while varying). As in images at higher frequencies, the intensity distribution in the image reveals a near-equatorial ``pancake'' distribution of electrons (with equatorial pitch angles close to 90$^\circ$) plus high-latitude lobes which require a component with a more isotropic distribution of pitch angles near L = 2. 

\begin{figure}[!ht]
\centering
\resizebox{1\hsize}{!}{\includegraphics{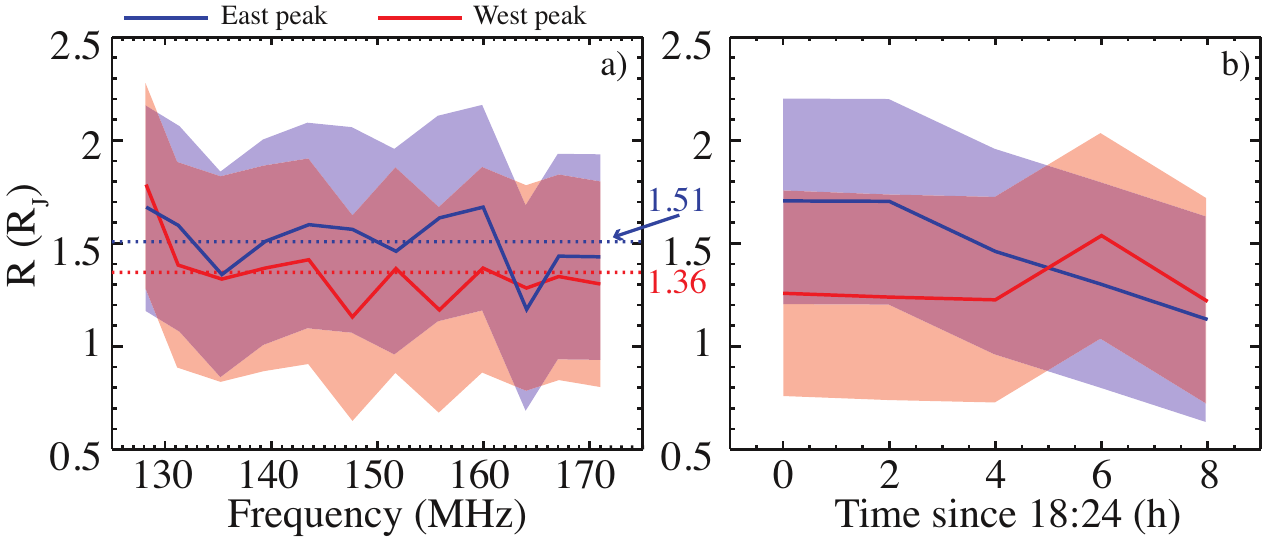}}
\caption{Radial distances of the east and west emission peaks as a function of frequency in time-integrated images (left) and as a function of time in frequency-integrated images (right). The range of variation of the peak position with frequency is [1.43 -- 1.67] $R_\text{J}$ for the east peak and [1.30 -- 1.78] $R_\text{J}$ for the west peak. With time, it is [1.13 -- 1.70] $R_\text{J}$ for the east peak and [1.22 -- 1.54] $R_\text{J}$ for the west peak. Shaded surfaces represent an uncertainty of $\pm1$ $R_\text{J}$.}

\label{fig9}
\end{figure}

We have measured the position of the ``east'' and ``west'' emission peaks as a function of frequency and time in (respectively) time-integrated and frequency-integrated images. The results are displayed in Fig. \ref{fig9} : in panels a) and b) an offset is measured between the average radial distance of the east maximum (1.51 $R_\text{J}$) and that of the west maximum (1.36 $R_\text{J}$). The accuracy of this measurement is limited to the size of the synthesized PSF for each of the reconstructed images. Although the determination of the peak flux is precise to the pixel level, we estimate the global uncertainty on the true position of the peaks to be $\sim$0.5 $R_\text{J}$ (as depicted by the error bars). Even with the lack of precision in our measurements, such east--west asymmetry was already observed (e.g. \citet{Dulk_1997} and \citet{SantosCosta_2009} at 5 GHz). It reveals the local time (dawn--dusk) asymmetry of the inner Jovian magnetosphere, also visible in the radial distance of the Io plasma torus and attributed to the presence of an east--west electric field \citep[see][and references therein]{Brice_1973,Smyth_1998,Kita_2013}. The time variations displayed in panels c) and d) of Fig. \ref{fig9}, measured at a few time steps, are consistent with radial excursions measured at higher frequencies \citep[$\sim$0.25 $R_\text{J}$ from 1.45 to 1.7 $R_\text{J}$ in][]{Dulk_1997}. Those are due to the longitudinal asymmetries of Jupiter's internal magnetic field that cause the average distance of the radiation belts to be dependent on the longitude, combined with projection effects on the sky at various phases of the planetary rotation. More accurate measurements are required to investigate this effect at low frequencies.

\begin{figure}[htpb]
\centering
\resizebox{0.75\hsize}{!}{\includegraphics{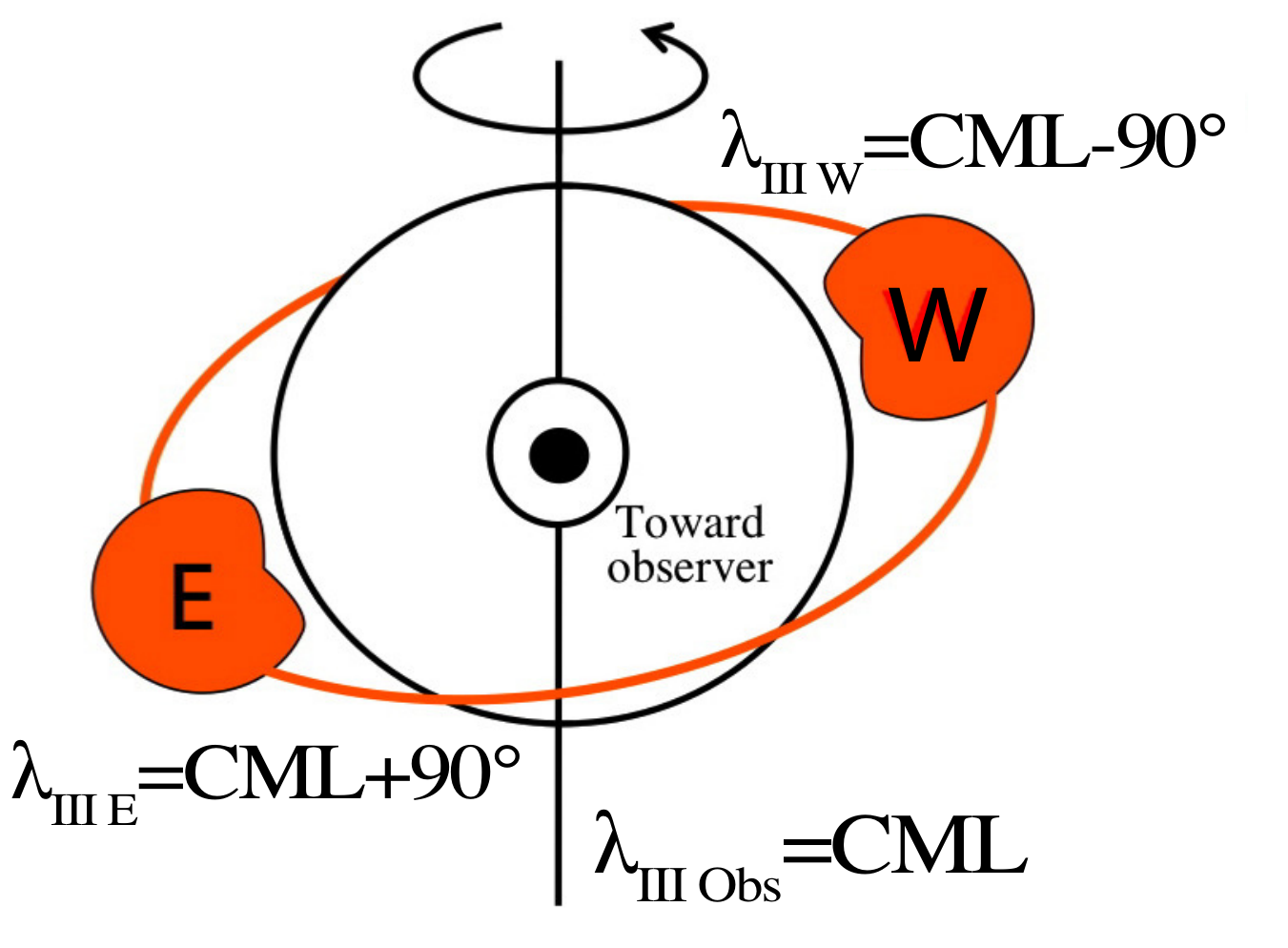}}
\caption{Sketch of the relationship between observer's CML and System III longitude of the east and west sides of a synchrotron image of Jupiter.}
\label{fig10a}
\end{figure}

\begin{figure}[htpb]
\resizebox{\hsize}{!}{\includegraphics{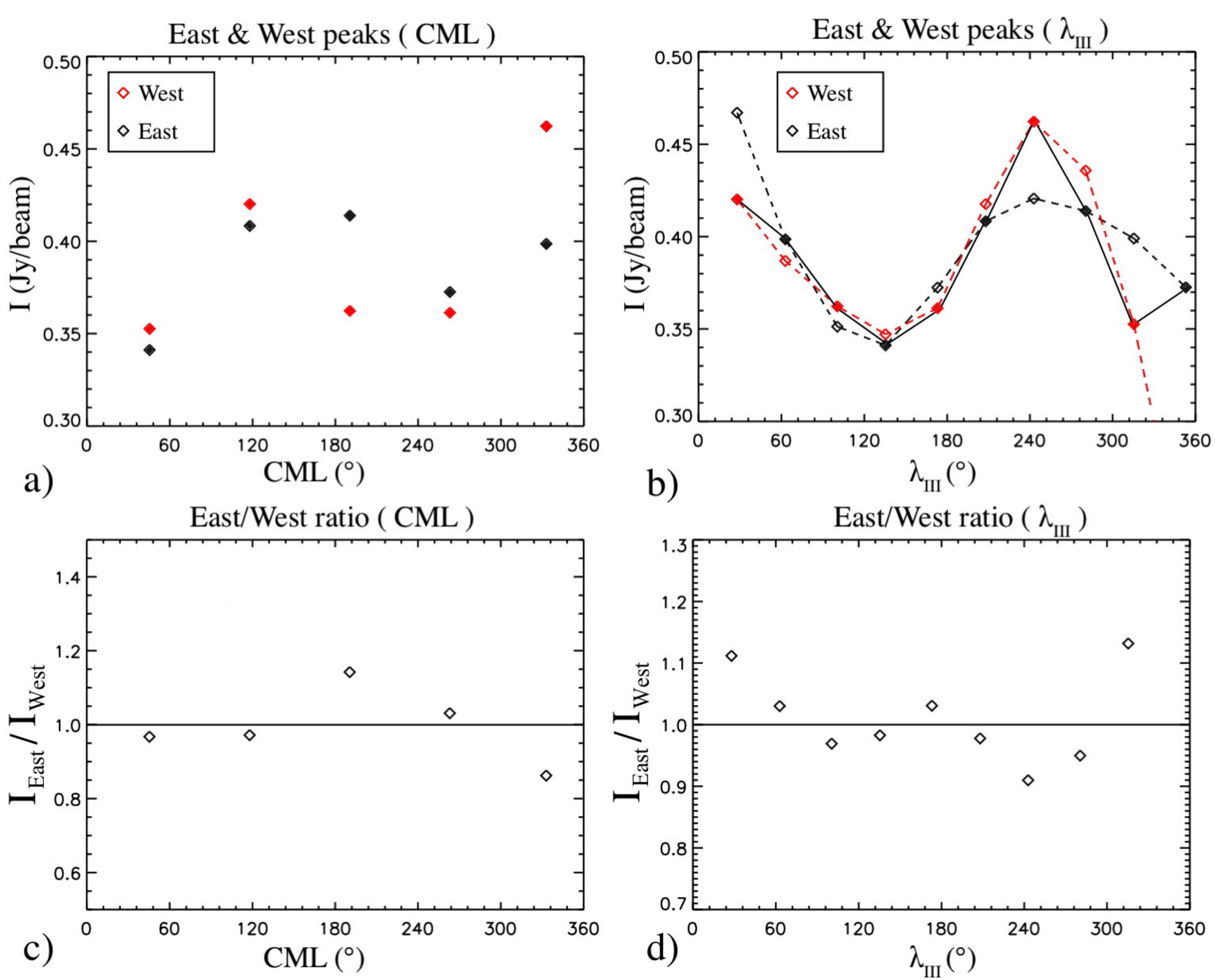}}
\caption{a) east and west peak intensities in each frequency-averaged image, as a function of the CML at the middle of the 2 hour interval corresponding to each image. b) east and west peak intensities -- as a proxy of the peak emissivity -- as a function of $\lambda_\text{III}$, derived from a) following the sketch of Fig. \ref{fig10a}. Measured values are filled diamonds connected by the solid line. Open diamonds are interpolated from a spline adjustment of east and west measurements separately. Dashed lines are the two resulting independent determinations of the peak emissivity profiles versus $\lambda_\text{III}$. c,d) east-to-west peak intensity ratios deduced from panels a) and b).}
\label{fig10b}
\end{figure}

Panel a) of Fig. \ref{fig10b} displays the peak intensity (in Jy/beam) measured on the east and west sides of Jupiter in each of the 5 frequency-averaged images, as a function of the CML at the middle of the 2 hour interval corresponding to each image. Following \citet{Dulk_1999b, Dulk_1999a} and as illustrated in Fig. \ref{fig10a}, although the observed emission from any point of the image results from integration along the line of sight through the optically thin radiation belts, the main contribution to the intensity observed at a given CML from the east side originates from a ``source" at System III longitude $\lambda_\text{III}$ = CML+90$^\circ$. Conversely, intensity observed on the west side originates from a ``source" at $\lambda_\text{III}$ = CML$-90^\circ$. Assuming that source characteristics (i.e., synchrotron emissivity at any point of the radiation belts) do not vary at timescales shorter than Jupiter's rotation period and that asymmetries in the magnetic field between the east and west sides can be ignored, it is possible by shifting by $\pm$90$^\circ$ the observed points on Fig. \ref{fig10b} a) to deduce a profile of the peak emissivity as a function of longitude, displayed on Fig. \ref{fig10b} b) as a solid line. To test the consistency of the above transformations from CML to $\lambda_\text{III}$, we adjusted separately the east and west measurements by a spline function and inferred intermediate values at each longitude where a measurement exists on the other side of Jupiter (open diamonds), resulting in pairs of values (derived from east and west peaks) at each longitude where one actual measurement exists. The two (dashed) profiles deduced from east and west peak intensities display similar overall variations. 

A broad hot spot is observed around $\lambda_\text{III}$ = 230$^\circ \pm 25^\circ$, that was already noted in previous observations at higher frequencies \citep{Branson_1968,Conway_1972}, and was suggested to be caused by the geometry of Jupiter's magnetic field configuration \citep{dePater_1980,dePater_1981}. east-to-west peak intensity ratios deduced from panels a) and b) of Fig. \ref{fig10b} are plotted in panels c) and d). The east/west ratio as a function of CML is reminiscent -- albeit with a lower amplitude -- of that measured at higher frequencies \citep[e.g.][]{dePater_1997,Kloosterman_2005,SantosCosta_2009}. 

The amplitude of the emission in our work is much lower, probably due to a combination of the long integration time (2 hr), the lower angular resolution in our images, and perhaps the lower frequency content of the source. As shown in \cite{Kloosterman_2005}, the detailed curves of the east/west ratio as a function of CML depend on the declination of the Earth relative to Jupiter ($D_\text{E}$). $D_\text{E}$ was different in each case ($-$3.3$^\circ$ for \cite{Leblanc_1997}, 0.07$^\circ$--0.34$^\circ$ for \cite{SantosCosta_2009}, and +3.29$^\circ$ in our observation), so that these comparisons are necessarily preliminary. 

Finally, our spectral measurements of Fig. \ref{fig8} are $\sim$35\% lower than earlier measurements from 1998 at the same frequency \citep{dePater_2003c}, i.e., marginally compatible with them taking into account our rather high estimated error bar ($\sim$30\%) on the LOFAR flux density measurements. But they are significantly lower than the model fit to the VLA measurement from 1994 \citep{dePater_2003c}. This suggests a possible turnover of the spectrum below $\sim$300 MHz and/or time variations of the spectral flux density overall (such as shown by the 1998 vs. 1994 data in Fig. \ref{fig7}), or just at low frequencies.

\section{Discussion and conclusion}

Synchrotron emission is a well-understood process. Observations allow us to probe the energetic electron population in the inner magnetosphere. At low frequencies, the thermal component is negligible, so that the emitted power proportional to $E^2 \times B^2$ at a peak frequency proportional to $E^2 \times B$ provides information about the lower energy part of this electron population. Resolved radio maps with good angular resolution provide important constraints (such as radial electron flux profile \& pitch angle distribution) to the physical radiation belts models, themselves built on models of electron acceleration and transport, pitch angle scattering, inward diffusion, effect of satellites, interaction with dust, losses, \dots  \citep{dePater_1981,dePater_1997,SantosCosta_2001b,Bolton_2004}. They also allow 3D reconstruction of the Jovian magnetic field topology close to the planet (thus sensitive to multipolar terms) by tomography \citep{Sault_1997,Leblanc_1997, dePater_1998}. Repeated observations permit us to characterize and study time variations that do exist at short timescales \citep{SantosCosta_2009, Tsuchiya_2011} or long timescales \citep{dePater_1989} which can be related to events such as the quick response and slow recovery after the impact of comet Shoemaker-Levy 9 \citep{dePater_1995, Millan_1998, Brecht_2001}, effects of asteroid impacts \citep{SantosCosta_2011}, Solar activity \citep{Kita_2013} or Solar wind fluctuations \citep{Bolton_1989,SantosCosta_2008}. The latter are still poorly understood.

LOFAR proves to be a powerful and flexible planetary imager, providing images complementary to the VLA at a spatial resolution only 4 times lower than the typical resolution at the VLA, but at frequencies >10 times lower, due to its long baselines. We have obtained here the first resolved images well below 300 MHz. Although still not perfect (LOFAR was still at commissioning stage) the image of Fig. \ref{fig6} roughly agrees with maps at higher frequencies; the shape of the emission confirms that two electrons populations coexist: a pancake and an isotropic one. The latter produces emission at high latitudes (near electron mirror points). We have characterized east--west or longitudinal asymmetries. Although the uncertainty of the LOFAR flux density value is high ($\sim$30 \%), The disk-integrated data points are stay marginally compatible with previous observations, suggesting the possible existence of a spectral turnover below 300 MHz and/or time variations of the spectrum.

LOFAR is now fully running. Further observations can be done with 24 core and 14 (possibly 16) remote stations (compared to 20 core and 9 remote stations in the paper) which will improve significantly the sensitivity and the angular resolution by the increase of long baselines (216 in this paper compared to 427 with 14 remote stations). Along with advanced hardware and software applied to the data, the set of 12 international (European) stations brings up the maximal baseline to 1500 km (instead of $\sim$100 km in the paper).
The Low Band of LOFAR will permit imaging of the Jovian synchrotron emission down to 40 MHz (upper limit of the decameter emission) and even less taking into account predictable absences of DAM emission \citep{Cecconi_2012}, bringing the first very low frequency images of Jupiter's radiation belts, diagnosing very low energy electrons and weak magnetic fields. Along with new LOFAR observations, joint synchrotron emission modeling is necessary.

 Another campaign was conducted on 19--20 Feb. 2013 with LOFAR LBA \& HBA and simultaneous Westerbork Synthesis Ratio Telescope observations at higher frequencies, ensuring together a frequency coverage from 50 MHz to 5 GHz ($\lambda$ = 6 m to 6 cm). It will allow us to address spectral variations and the search for a low-frequency turnover. Further studies will also rely on the analysis of the polarization of the emission \citep[long known to be dominantly linear, e.g.,][]{Radhakrishnan_1960,dePater_1980}. 
Advanced imaging methods such as sparse image reconstruction \citep{Garsden_2015,Girard_2015} of the extended emission may improve the quality of snapshot images to better constrain the shape of the belts in smaller CML integration windows.
Finally, synchrotron observations in the context of the JUNO mission around Jupiter will also be of high interest, as JUNO will provide \textit{in situ} particle measurements and a very accurate model of the Jovian magnetic field \citep{Bagenal_2014}.

\begin{acknowledgements}
We acknowledge the financial support from the UnivEarthS Labex program of Sorbonne Paris Cit\'e (ANR-10-LABX-0023 and ANR-11-IDEX-0005-02) and from the European Research Council grant SparseAstro (ERC-228261).
LOFAR, the Low Frequency Array designed and constructed by ASTRON, has facilities in several countries, that are owned by various parties (each with their own funding sources), and that are collectively operated by the International LOFAR Telescope (ILT) foundation under a joint scientific policy. The authors thank Roberto Pizzo (ASTRON, Dwingeloo) and Jean-Mathias Grie\ss meier (LPC2E, Orl\' eans) for their assistance with the observations. James M. Anderson (co-author) for a thorough final inspection of typos and notation convention.
\end{acknowledgements}
\FloatBarrier

\bibliographystyle{aa}
\bibliography{references}


\begin{appendix}

\section{Coordinate transformations and Jupiter tracking}
\subsection{Phase center correction}

We defined the rotation $\mathcal{R}_w$ (resp. $\mathcal{R}_u$), the rotation of angle $-\alpha_0$ (resp. $\delta_0$) around the axis $w$ (resp. $u$) in the ($u$,$v$,$w$) space. The ($\alpha_0$,$\delta_0$) defines the equatorial coordinates of the phase center, which was maintained constant during the observation.
We want to apply the angular transformation from ($\alpha_0$,$\delta_0$) to ($\alpha_t$,$\delta_t$) where $\alpha_t$ and $\delta_t$ are the time-dependent center coordinates of the Jupiter disk during the observation. We used the ephemeris from the Institut de M\'ecanique C\'eleste et de Calcul des \'Eph\'em\'erides (IMCCE) to locate the center of the disk in equatorial coordinates. The correction was performed at a 5 minute rate. Given the orientation of the declination and right ascension axes, we can define two rotation matrices around the axis $w$ and the axis $u$ as follows:

\begin{equation}
\mathcal{R}_w(-\alpha_t) = 
\begin{pmatrix}
\cos \alpha_t & \sin \alpha_t & 0 \\
-\sin \alpha_t & \cos \alpha_t & 0 \\
 0 & 0 & 1 
\end{pmatrix}
\label{eq:rotw}
\end{equation}

\begin{equation}
\mathcal{R}_u(\delta_t) = 
\begin{pmatrix}
1 & 0 & 0\\
0 &\cos \delta_t & -\sin \delta_t  \\
0 & \sin \delta_t & \cos \delta_t
\end{pmatrix}
\label{eq:rotu}
\end{equation}


The operator $\mathcal{T}$ to transform the frame toward the direction of Jupiter at time $t$ is therefore:

\begin{equation}
\mathcal{T}_t= \mathcal{R}_u(\delta_t) \mathcal{R}_w(-\alpha_t)= 
\begin{pmatrix}
\cos \alpha_t & \sin \alpha_t \cos \delta_t & -\sin \alpha_t \sin \delta_t\\
-\sin \alpha_t &\cos \alpha_t \cos \delta_t& -\cos \alpha_t \sin \delta_t \\
0 &  \sin \delta_t & \cos \delta_t
\end{pmatrix}
\label{eq:matt}
\end{equation}
In addition, it is required to apply a phase correction to the complex visibility data as the plane wave coming from direction $\vec{u}_0$ should now come in phase from direction $\vec{u}_t$. This factor is expressed as a function of the transformation:

\begin{equation}
  \underline{\phi_\text{cor}}(\lambda,t)=\exp{(j\frac{2\pi}{\lambda} ([w_0-w_t]^\top \mathcal{T}_t).[u_t,v_t,w_t]^\top)}
\end{equation}

where $w_0$ and $w_t$ are the third column of matrix Eq. \ref{eq:matt} with the corresponding indices.

\subsection{Intrinsic rotation correction}
Once the previous phase and axis corrections have been performed, we need to apply a correction on the ($u$,$v$) axes to follow the intrinsic oscillation of the radiation belts.
 By applying a time-dependent rotation of angle $\beta_\text{m}(t)$ (with $\beta_\text{m}(t) = -111.6^{\circ} \pm 9.6^{\circ}$, counting from the increasing declination axis), the mean direction of the apparent magnetic equator on the sky, around the axis defined by $\vec{w}_J$ by the following transformation:

\begin{equation}
\mathcal{R}_w(\beta_\text{m(t))} = 
\begin{pmatrix}
\cos \beta_\text{m}(t) & -\sin \beta_\text{m}(t) & 0 \\
\sin \beta_\text{m}(t) & \cos \beta_\text{m}(t) & 0 \\
 0 & 0 & 1 
\end{pmatrix}
\end{equation}
\label{eq:wobble}
\end{appendix}

In first approximation, no phase correction is necessary after applying the rotation of Eq. \ref{eq:wobble} on the ($u$,$v$,$w$) coordinates.

\end{document}